\documentclass[final,preprint]{elsarticle}

\usepackage{epsfig}
\usepackage{bm}
\usepackage{amsmath,amssymb,amsthm}
\usepackage{graphicx,color}
\usepackage{psfrag}
\usepackage{bm}

\def\s2{\sigma^2}

\begin{document}
\begin{frontmatter}

\title{Zitterbewegung in monolayer silicene in a magnetic field}

\author[3c]{E. Romera}
\author[1a]{J. B. Rold\'an}
\author[2b]{F. de los Santos}
\address[3c]{ Departamento de F\'isica At\'omica, Molecular y Nuclear and
  Instituto Carlos I de F{\'\i}sica Te\'orica y Computacional, Universidad de
  Granada, Fuentenueva s/n, 18071 Granada, Spain}
\address[1a]{ Departamento de Electr\'onica y Tecnolog\'{\i}a de Computadores \& CITIC,
  Universidad de Granada. Fuentenueva s/n, 18071 Granada, Spain.}
\address[2b]{ Departamento de Electromagnetismo y F{\'\i}sica de la
Materia, and Instituto Carlos I de F{\'\i}sica Te\'orica y
Computacional, Universidad de Granada, Fuentenueva s/n, 18071 Granada,
Spain}

\date{\today}

\begin{abstract}
We study the Zitterbewegung in monolayer silicene under a perpendicular magnetic field. 
Using an effective Hamiltonian, we have investigated the autocorrelation function and the density currents in this 
material. Moreover, we have analyzed  other types of periodicities of the 
system (classical and revival times). 
Finally, the above results are compared with their counterparts in two other
monolayer materials subject to a magnetic field: graphene and MoS$_2$ 

\end{abstract}
\end{frontmatter}



{\em Introduction}.
Silicene is a monolayer of Silicon atoms theoretically predicted some years ago
\cite{takeda,guzman94}  and experimentally observed 
in 2010 \cite{vogt12,Aufray10,lalmi10,fleurence12,padova}. It has a honeycomb structure similar to that
of graphene and its low-energy electronic properties can also be described
by an effective Dirac-type Hamiltonian for each of the two inequivalent corners of the Brillouin
zone \cite{tahir2013,ezawa1,ezawa2,drummond,ezawa3,ezawa4}. Silicene has a band gap of 1.55 meV due to a relative large
spin-orbit coupling which can be experimentally controlled  by applying a
perpendicular electric field.  Interesting physical
properties have been investigated in this material, as a quantum valley Hall effect or topological phase transitions.

On the other hand, Zitterbewegung (ZB) appears in Dirac particles as a rapid trembling motion
around their otherwise rectilinear average trajectory, as a
consequence of the interference between eigenstates with positive and negative
eigenvalues. Recently, there have been many theoretical studies of ZB, not only
in relativistic systems but also in condensed matter systems  (see the review by Rusin and
Zawadski \cite{review} and references therein). ZB has not been
experimentally observed yet due to its large frequency and small amplitude,
but it has been simulated experimentally by means of trapped ions (adjusting
some parameters of the Dirac equation) by  Gerritsma et
al. \cite{Gerristma10}. ZB has been studied in different types of monolayer and bilayer graphene devices
\cite{review,17,41,46,47,48,51,54,55,56}.

In this Letter,  we shall  describe ZB in silicene in a perpendicular
magnetic field. Additionally, we shall show that there is 
a regeneration of the ZB amplitude for appropriate wave packets's initial conditions. Next, we
shall introduce the effective Hamiltonian for silicene and then study the
ZB and revivals. Finally, we shall compare the results with two other
monolayer materials, namely, graphene and MoS$_2$.


{\em Model Hamiltonian}. 
Let us consider a monolayer silicene  in a perpendicular magnetic
field $B$  with an effective Hamiltonian given by \cite{tahir2013}
\begin{equation}
H_s^{\tau}=v_F (\sigma_x  p_x-\tau \sigma_y  p_y )-\tau
s \Delta \sigma_z,
\label{hamiltoniano}
\end{equation}
where $\tau=\pm 1$ corresponds respectively to the inequivalent corners ${\bm K}$ and
${\bm K}^{\prime}$ of the Brillouin zone, ${\bm p}$ stands for the momentum,
${\bm \sigma}$ is the Pauli matrices vector, $v_F=5\times 10^5$ m/s is the Fermi
velocity of the Dirac fermions, with up and down spin values being
represented by $s=\pm 1$, and $\Delta$ is the band gap  induced by
intrinsic spin-orbit interaction, which provides a mass to the Dirac fermions. 
Using the Landau gauge, ${\bm A}=(0, Bx,0)$, the eigenvalues can be
written as \cite{tahir2013}
\begin{equation}
E^{\tau}_{s n \gamma}=\gamma \sqrt{n\epsilon^2 + \Delta^2},
\label{energia}
\end{equation}
with $\gamma=\pm1$ accounting for the valence and conduction bands and $\epsilon=\sqrt{2}\hbar
v_F /l_B\approx 0.1794 \sqrt{B(T)}$ eV. The corresponding
eigenfunctions  at the ${\bm K}$ point  are given by
\begin{eqnarray}
\Phi_{s n \gamma}({\bf r}) = & e^{ik_y y} \left( \begin{array}{cl}
                    \alpha_{\gamma, n} \langle x|n-1\rangle \\
                    \beta_{ n}\langle x|n\rangle \\
                    \end{array}
                    \right), n\geq 1, \\
 & \left( \begin{array}{cl}
                    |0\rangle \\
                    0 \\
                    \end{array}
                    \right), n=0.
\label{spinor}
\end{eqnarray}
Here, $\alpha_{\gamma,n}=\Delta+\gamma \sqrt{\Delta^2+n\epsilon^2}$,  
$\beta_{n}=-i\sqrt{n}\epsilon$ and with normalization given by
$N_{\gamma,n}=\sqrt{|\alpha_{\gamma,n}|^2+|\beta_{n}|^2}$. The eigenfunctions
at the ${\bm K^{\prime}}$ ($\tau=-1$) point are obtained by exchanging electron and
hole components in the above case. We shall take $\Delta=4$ meV.
The eigenvalues and eigenvectors do not depend on
the values of $s$ within this level of approximation, so we do not take
into account the spin of the electron in this study

{\em Wave packet temporal evolution}.
The evolution of a wave packet  for a time independent Hamiltonian is
given by (in units such that $\hbar=1$)
\begin{equation}
\Psi(t)=\displaystyle\sum_{n=0}^{\infty} a_n u_n e^{-i E_n t}
\label{evo}
\end{equation}
$u_n$ and $E_n$ being the eigenfunctions and eigenvalues, respectively,
and $a_n=\langle u_n,\Psi\rangle$. If we consider an initial
wave packet localized around  the energy $E_{n_0}$ of the spectrum
$E_n$, then a Taylor expansion of the energy can be written as
\begin{equation}
E_n\approx
E_{n_0}+E_{n_0}^{\prime}(n-n_0)+\frac{E_{n_0}^{\prime\prime}}{2}
(n-n_0)^2+...
\label{expan} 
\end{equation}
and, consequently, 
\begin{eqnarray}
\exp{\left(-i E_n t\right)}= \hspace{6cm}\notag\\  \exp{\left(-i\left[E_{n_0}  + E_{n_0}^{\prime}(n-n_0) +
\frac{E_{n_0}^{\prime\prime}}{2}(n-n_0)^2+\cdots\right]t\right)} \nonumber\\
\hspace{-2cm} \exp{\left(-iE_{n_0} t - \frac{2\pi i(n-n_0) t}{  T_{\rm Cl}} -\frac{2\pi i (n-n_0)^2 t}{ 
T_{\rm R}} + \cdots\right)},\nonumber\\
\label{approx}
\end{eqnarray}
where each term defines an important characteristic time scale.
In particular, the so-called revival time $T_{\rm  R}\equiv {4\pi}/{|E_{n_0}^{\prime\prime}|}$ 
is the time the wave packet needs to return to a shape that is approximately the
same as the initial one in the temporal evolution,
and the classical period, $T_{\rm Cl}\equiv {2\pi}/{|E_{n_0}^{\prime}|}$, is 
the time over which the wave packet follows the expected classical trajectory 
(see \cite{54} and references therein for more details).

Furthermore, we shall study ZB  in the wave packet evolution. To this purpose, we have elected as the initial 
state a superposition of two wave packets with both positive and negative energy levels,
which we denote respectively by $E_n^{(+)}$ and $E_n^{(-)}$,  and with associated eigenfunctions 
$u_n^{(+)}$ and $u_n^{(-)}$. To be concrete,
\begin{equation}
 \psi=\frac{1}{\sqrt{2}} \left(\psi_+ + \psi_-\right)
\end{equation}
with
\begin{eqnarray}
\psi_+ &=& \displaystyle\sum_{n=0}^{\infty} c_n^{(+1)} \Phi_{s n 1}(x,y), \nonumber \\
\psi_- &=& \displaystyle\sum_{n=0}^{\infty} c_n^{(-1)} \Phi_{s n -1}(x,y).
\label{wp2}
\end{eqnarray}
Under these conditions, ZB will appear as an additional periodicity in 
the wave packet evolution given by $T_{\rm  ZB}={\pi}/{E_{n_0}}$ \cite{54}.  
Lastly, we mention that the choice of our initial wave packet is one
used in many theoretical works, but for experimental purposes it should be taken into account 
that it has recently been described how to create more realistic wave packets by illuminating 
Carbon nanotubes with short laser pulses \cite{1Rusin2014}.

{\em Results}.
First, we shall consider the ZB in the wave packet evolution in 
monolayer silicene. Both packets (see Eq. \ref{wp2}) are centered 
around the same value of $n_0=11$, and with coefficients Gaussianly distributed as 

\begin{equation}
c_n^{(\pm 1)}= c_{n}=
\left(\frac{1}{\pi\sigma^{1/2}}\right)^{1/2}
\exp{\left(-\frac{(n-n_0)^2}{2\sigma}\right)}.
\label{gaussian_cn}
\end{equation}
Other parameter values are $B=1$T,  $\sigma=3$
and $s=1$. The subsequent time evolution of this 
initial state is then monitored by the computation of the 
autocorrelation function, $A(t)$, customarily defined as 
\begin{eqnarray}
A(t) &= &\int dx dy \Psi^{*}(x,y,0)\Psi(x,y,t).
\end{eqnarray}
Notice that a value of $|A(t)|^2$ close to unity implies that the initial and the
time-evolved wave packets must have a significant overlap.
We shall make use of this quantity to identify the several types of periodicity
in the temporal development of the wave packet, namely, ZB, classical
motion and revivals. These can be calculated
 analytically  using Eq. (\ref{energia}) and read
\[
T_{\rm ZB}=\frac{\pi}{\gamma\sqrt{\Delta^2 + \epsilon^2 n}},\quad
T_{\rm Cl}=\frac{4\pi\sqrt{\Delta^2 + \epsilon^2 n}}{\epsilon^2 \gamma},
\]
\[
T_{\rm R}=\frac{16\pi(\Delta^2 + \epsilon^2 n)^{3/2}}{\epsilon^4 \gamma}.
\]
\begin{figure}[t]
\includegraphics[width=\linewidth]{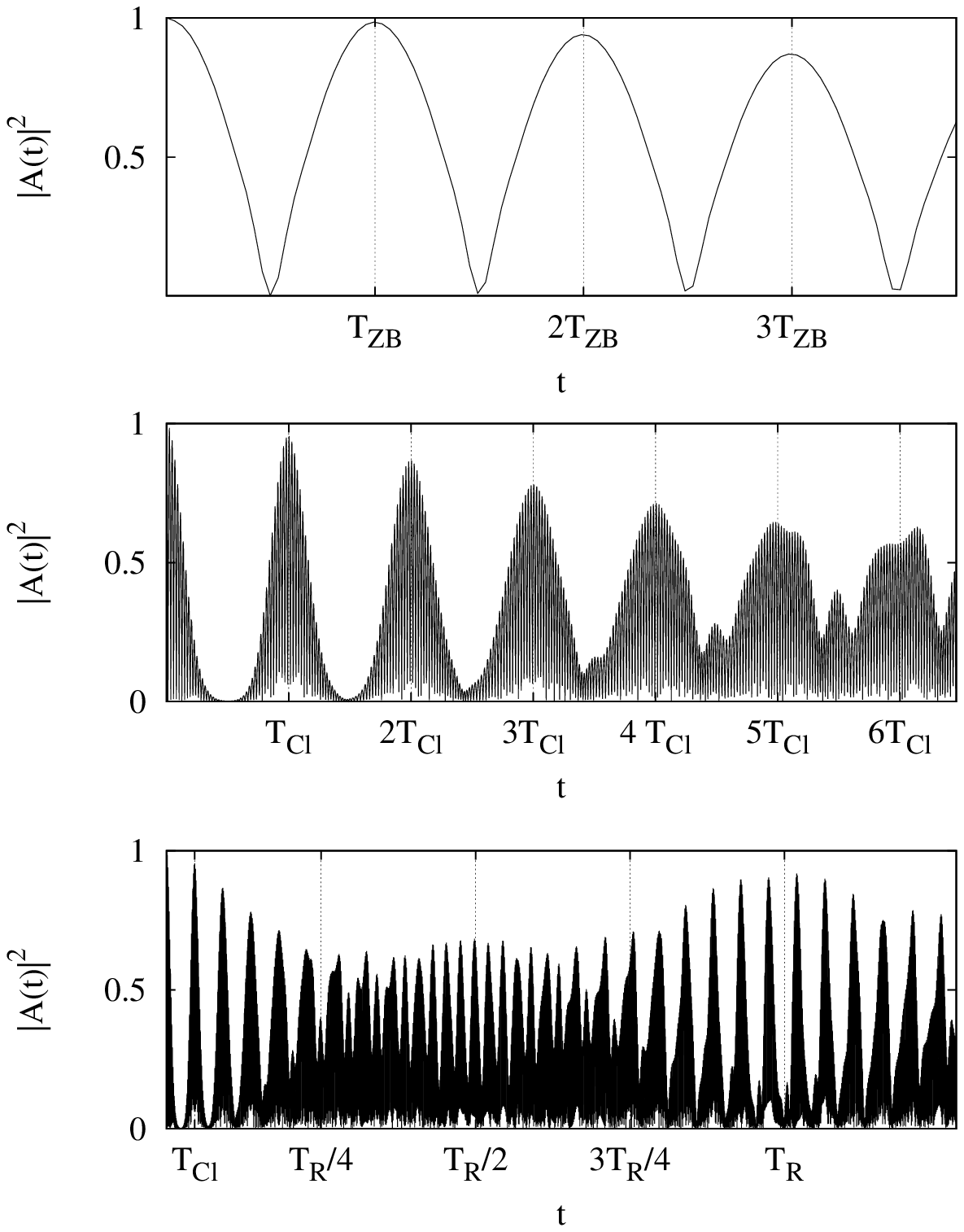}
\caption{
Time dependence of the autocorrelation function $|A(t)|^2$ in silicene for
$B = 1$ T, $n_0 = 11$, $\sigma = 3$, a band gap  $\Delta=4$ meV,  and
$s=1$. (top panel) ZB of electrons with $T_{\rm ZB} \simeq 
3.47$ 
fs
(central panel) First classical periods of motion with $T_{\rm
  Cl}\simeq 0.15$ ps. (bottom panel) Long-time dependence with $T_{\rm R}\simeq 3.3$ ps.
The Zitterwebegung,  classical periods and the main fractional revivals are indicated by vertical
dotted lines.
}
\end{figure}
Figure 1 shows the time dependence of 
$|A(t)|^2$ at the three different time scales. The top panel  represents the ZB
of electrons with $T_{\rm ZB} \simeq 0.47$ fs; in the middle panel  the first 
classical periods of motion can be appreciated, with $T_{\rm Cl} \simeq 0.15$ ps;
the bottom panel  displays the long time revival behavior with $T_{\rm R} \simeq 3.3$ ps
along with the several fractional revivals occurring at intervening times.
The main fractional revivals are indicated by vertical, dotted lines, where 
$T_{\rm R}$ is calculated from its analytical expression $T_{\rm R}=4\pi/|E''_{n_0}|$.

Additionally, we have also studied the time dependence of electronic currents 
and how they are affected by the three above-mentioned types of oscillatory motion.
In particular, since the electron velocity operators are given by 
$v_{j}=i[H,r_{j}]/=v_F\sigma_{j}$, with $j=x,y$, it is straightforward to obtain $j_x(t)=0$ and 
\begin{eqnarray}
 j_y(t)=2e v_F\displaystyle\sum_{n=0}^{\infty} c_n c_{n-1}
 \xi_n 2\cos(E_nt)\cos(E_{n-1}t).
\end{eqnarray}
with $\xi_n=[(\alpha_{+,n}/N_{+,n})+(\alpha_{-,n}/N_{+,n})]/\beta_{n-1}$.
 
Figure 2 shows the influence of ZB on the current in the femtosecond
scale (top panel);
ZB oscillations can be clearly seen superimposed on the classical ones at medium times
(middle panel); revivals of the electric current can be recognized at $T_{\rm R}$ and fractional 
revivals at, for instance, $T_{\rm R}/2$ and $3T_{\rm R}/4$, in the
picosecond scale (bottom panel).

Next, we extend the previous study to MoS$_{2}$ and graphene. The Hamiltonian for the 
former is identical to that of monolayer silicene except for its smaller gap
$\Delta=0.775$ eV, Fermi velocity $v_F=85000$ m/s, and
consequently $\epsilon \simeq 30.5 10^{-3} \sqrt{B}$. In the case of graphene, 
within the tight binding approximation and near the vicinity of the ${\bm K}$ point, 
one of the two inequivalent corners of the Brillouin zone, the Hamiltonian
reads as in Eq. (\ref{hamiltoniano}) but now
with $\Delta=0$ and the Fermi velocity $10^6$ m/s.

In Figure 3 results are compared for silicene, MoS$_2$ and graphene. The top panel shows that the ZB time is
almost insensitive to the intensity of the magnetic field for MoS$_2$
(red, long dashed line), whilst it rapidly decays for 
silicene (brown, short dashed line) and graphene (blue, solid line).
The classical periods are compared in the central panel, 
where it can be appreciated that MoS$_2$ behaves differently in that its classical time is almost two orders of magnitude higher 
than those of silicene and graphene for any $B$. The same trend repeats in the bottom panel, showing that the revival times
of silicene and graphene are similar, but that of MoS$_2$ is approximately two orders of magnitude higher at large $B$ and
four or more orders of magnitude higher at low $B$. 
As $B$ approaches zero, ZB times for MoS$_2$, silicene and graphene converge to finite values that are of the same order of 
magnitude. The revival and classical times diverge when $B$ approaches zero in the three cases.

\begin{figure}[t]
\includegraphics[width=\linewidth]{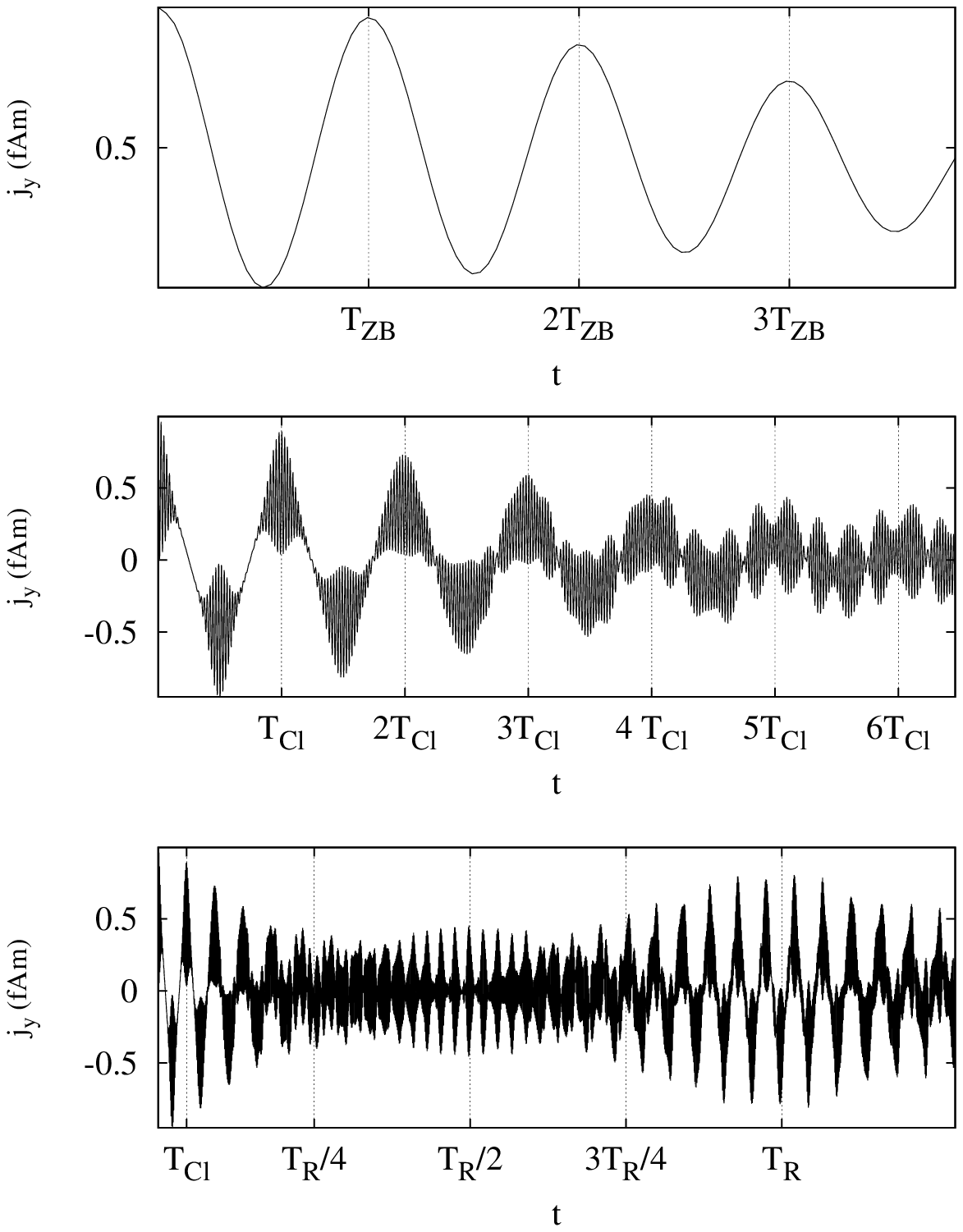}
\caption{
Time dependence of electric currents $j_y$  in silicene for
$B = 1$ T, $n_0 = 11$, $\sigma = 3$, a band gap  $\Delta=4$meV and
$s=1$. (Top panel) ZB of electrons with $T_{\rm ZB} \simeq 3.47$  fs.
(Middle panel) First classical periods of motion with $T_{\rm
  CL}\simeq 0.15$ ps. (Bottom panel) Long-time dependence with $T_{\rm R}\simeq 3.3$ ps.
The Zitterwebegung,  classical periods and the main fractional revivals are indicated by vertical
dotted lines.
}
\end{figure}

\begin{figure}
\includegraphics[width=8cm]{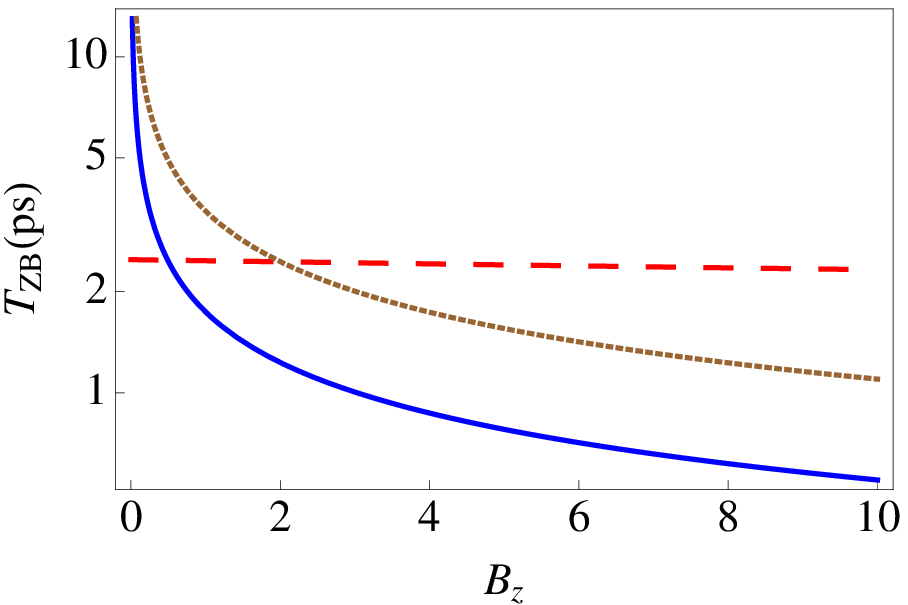}
\includegraphics[width=8cm]{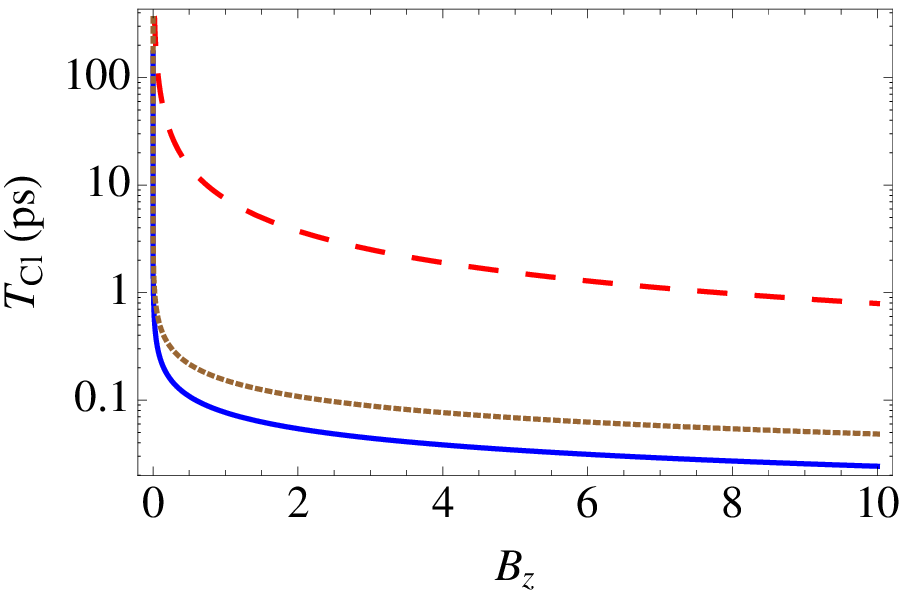}
\includegraphics[width=8cm]{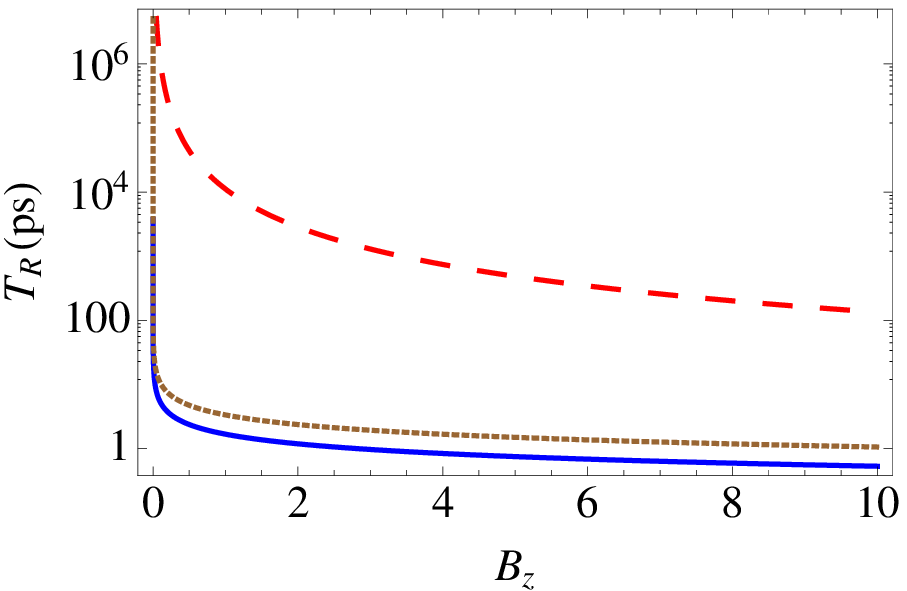}
\caption{ZB (Top panel), classical (middle panel), and revival times (bottom panel) for 
silicene (brown, short dashed  line), MoS$_2$ (red, long dashed line), and graphene (blue, solid line).}
\end{figure}

{\em Summary}.
We have studied the current oscillations due to ZB in silicene under a magnetic 
field, by considering wave packets with a Gaussian population of both positive 
and negative energy levels. Additionally, the ZB time is compared with other periodicities 
that arise when the wave packets are peaked around a sufficiently large quantum number,
namely the classical period and the revival times. As in a previous study \cite{54},
we find $T_{\rm ZB} < T_{\rm Cl}< T_{\rm R}$, with $T_{\rm ZB}$ in the femtosecond scale
and $T_{\rm Cl}, T_{\rm R}$ in the picosecond scale for a moderate intensity of the magnetic
field, $B=1$ T. A comparison with MoS$_2$ and graphene reveals that despite
these three materials share a similar hexagonal monolayer structure
the classical and revival time scales of MoS$_2$ are largely different from those
of silicene  and graphene. Interestingly, also MoS$_2$'s ZB times differ drastically
from silicene and graphene, showing that $T_{\rm ZB}$ for MoS$_2$ is almost
insensitive to the strength of the magnetic field. 
We conclude by mentioning that there is a great interest in the experimental observation of ZB and, 
in this regard, we point out a recent relevant proposal to observe this phenomenon by using the techniques 
of quantum optics \cite{RZ}.

{\em Acknowledments}.
Our work has been supported by the Spanish Projects No. MICINN FIS2009-08451, MICINN FIS2011-24149, CEI-BIOTIC Granada PV8  and Junta de Andaluc\'{\i}a P12.FQM.1861.


\begin{thebibliography}{99}

\bibitem{takeda} 
K. Takeda, K. Shiraishi, 
Phys. Rev. B 50 (1994) 14916.

\bibitem{guzman94}
G. G. Guzman-Verri, L. Lew Yan, 
Phys. Rev. B 76 (2007) 075131.

\bibitem{vogt12} 
P. Vogt et al., Phys. Rev. Lett.  108 (2012) 155501.

\bibitem{Aufray10} 
B. Augray, A. Kara, S. B. Vizzini, H. Oughaldou, 
C. L\'eAndri, B. Ealet, G. Le Lay,
App. Phys. Lett. 96 (2010) 183102.  

\bibitem{lalmi10} 
B. Lalmi, H. Oughaddou, H. Enriquez, A. Kara, S. B.   Vizzini, B. N. Ealet, B. Augray,
App. Phys. Letters 97 (2010) 223109.

\bibitem{fleurence12} 
A. Feurence, R. Friedlein, T. Ozaki, H. Kawai, Y. Wang, Y. Y. Takamura, 
Phys. Rev. Lett. 108 (2012) 245501.

\bibitem{padova} 
P. E. Padova et al.,
App. Phys. Lett. 96 (2010) 261905.

\bibitem{tahir2013} 
M. Tahir, U. Schwingenschl\"ogl, 
Sci. Rep. 3 (2013) 1075.

\bibitem{ezawa1} 
M. Ezawa, 
New J. Phys. 14 (2012) 033003.

\bibitem{ezawa2} 
M. Ezawa, 
J. Phys. Soc. Jpn. 81 (2012) 064705.

\bibitem{drummond} 
N. D. Drummond, V. Z\'olyomi, V. I. Fal'ko, 
Phys. Rev. B  85 (2012) 075423.

\bibitem{ezawa3} 
M. Ezawa, 
Phys. Rev. B 86 (2012) 161407 (R).

\bibitem{ezawa4} 
M. Ezawa, 
Phys. Lett. A 378 (2014) 1180. 


\bibitem{review}
W. Zawadzki, T. M. Rusin, 
J. Phys. Condens. Matter 23 (2011) 143201.

\bibitem{Gerristma10} 
R. Gerritsma, G. Kirchmair, F. Zahringer, E. Solano,
R. Blatt, C. F. Roos, 
Nature (London) 463 (2010) 68.

\bibitem{17} 
T. M. Rusin, W. Zawadzki,
Phys. Rev. B 76 (2007) 195439;
78 (2008) 125419; 80 (2009) 045416.

\bibitem{41}
J. Cserti and G. D\'avid, 
Phys. Rev. B 74 (2006) 172305,
Phys. Rev. B 82 (2010) 201405(R);
G. D\'avid and J. Cserti;
Phys. Rev. B 81 (2010) 121417(R).

\bibitem{46} 
M. I. Katsnelson, 
Europ. Phys. J. B 51 (2006) 157.

\bibitem{47} 
G. M. Maksimova, V. Y.  Demikhovskii, E. V. Frolova, 
Phys. Rev. B 78 (2008) 235321.

\bibitem{48} 
J. C. Mart\'inez, M. B. A. Jalil,  S. G. Tan,
Appl. Phys. Lett. 97 (2010) 062111.

\bibitem{51} 
J. Schliemann,  
New J. Phys. 10 (2008) 034024.

\bibitem{54} 
E. Romera, F. de los Santos,
Phys. Rev. B 80 (2009) 165416.

\bibitem{1Rusin2014} 
T. M. Rusin, W. Zawadzki,
J. Phys. Cond. Matter 26 (2014) 215301.

\bibitem{55} 
Y. X.  Wang, Z.  Yang, S. J. Xiong, 
Eur. Phys. Lett. 89 (2010) 17007.

\bibitem{56}  
T. Garc\'{\i}a, N. A. Cordero, E. Romera, 
Phys. Rev. B 89 (2014) 075416.
\bibitem{RZ}  T. Rusin and W. Zawadzki
arXiv: 1404.6187 (2014).

\end{thebibliography}
\end{document}